\documentclass[twocolumn,secnumarabic,amssymb, amsmath, nofootinbib,tightenlines,
nobibnotes, showpacs, superscriptaddress, aps, prl]{revtex4}
\usepackage{graphicx}

\def\Journal#1#2#3#4{{#1} {\bf #2}, #3 (#4)}
\def\FBSS{\rm Few Body Syst. Supp.}
\def\NIMA{{\rm Nucl. Inst. Meth.} A}
\def\NPA{{\rm Nucl. Phys.} A}
\def\PRP{\rm Phys. Rep.}
\def\PRC{{\rm Phys. Rev.} C}
\def\PRL{\rm Phys. Rev. Lett.}
\def\PLB{{\rm Phys. Lett.} B}

\def\reaccont{$^3$He$(e,e^{\prime}p)pn$}
\def\reactwobbu{$^3$He$(e,e^{\prime}p)d$}
\def\reacboth{$^3$He$(e,e^{\prime}p)$}
\def\reacx{$^3$He$(e,e^{\prime}p)X$}
\def\reacqe{$^3$He$(e,e^{\prime})$}
\def\etal{{\it et al.}}

\begin{document}
%\draft command makes pacs numbers print
%\draft

\title{
Measurement of the {\reaccont} reaction at high missing energies and momenta
}
\author{F.~Benmokhtar}
\affiliation{Rutgers, The State University of New Jersey, Piscataway, New Jersey 08854, USA}
\affiliation{Universit\'{e} des Sciences et de la Technologie, BP 32, El Alia, Bab Ezzouar, 16111 Alger, Alg\'{e}rie}
\author{M.~M.~Rvachev}
\affiliation{Massachusetts Institute of Technology, Cambridge, Massachusetts 02139, USA}
\author{E.~Penel-Nottaris}
\affiliation{Laboratoire de Physique Subatomique et de Cosmologie, F-38026 Grenoble, France}
\author{K.~A.~Aniol}
\affiliation{California State University, Los Angeles, Los Angeles, California 90032, USA}
\author{W.~Bertozzi}
\affiliation{Massachusetts Institute of Technology, Cambridge, Massachusetts 02139, USA}
\author{W.~U.~Boeglin}
\affiliation{Florida International University, Miami, Florida 33199, USA}
\author{F.~Butaru}
\affiliation{Laboratoire de Physique Subatomique et de Cosmologie, F-38026 Grenoble, France}
\author{J.~R.~Calarco}
\affiliation{University of New Hampshire, Durham, New Hampshire 03824, USA}
\author{Z.~Chai}
\affiliation{Massachusetts Institute of Technology, Cambridge, Massachusetts 02139, USA}
\author{C.~C.~Chang}
\affiliation{University of Maryland, College Park, Maryland 20742, USA}
\author{J.~-P.~Chen}
\affiliation{Thomas Jefferson National Accelerator Facility, Newport News, Virginia 23606, USA}
\author{E.~Chudakov}
\affiliation{Thomas Jefferson National Accelerator Facility, Newport News, Virginia 23606, USA}
\author{E.~Cisbani}
\affiliation{INFN, Sezione Sanit\'a and Istituto Superiore di Sanit\'a, Laboratorio di Fisica, I-00161 Rome, Italy}
\author{A.~Cochran}
\affiliation{Hampton University, Hampton, Virginia 23668, USA}
\author{J.~Cornejo}
\affiliation{California State University, Los Angeles, Los Angeles, California 90032, USA}
\author{S.~Dieterich}
\affiliation{Rutgers, The State University of New Jersey, Piscataway, New Jersey 08854, USA}
\author{P.~Djawotho}
\affiliation{College of William and Mary, Williamsburg, Virginia 23187, USA}
\author{W.~Duran}
\affiliation{California State University, Los Angeles, Los Angeles, California 90032, USA}
\author{M.~B.~Epstein}
\affiliation{California State University, Los Angeles, Los Angeles, California 90032, USA}
\author{J.~M.~Finn}
\affiliation{College of William and Mary, Williamsburg, Virginia 23187, USA}
\author{K.~G.~Fissum}
\affiliation{University of Lund, Box 118, SE-221 00 Lund, Sweden} 
\author{A.~Frahi-Amroun}
\affiliation{Universit\'{e} des Sciences et de la Technologie, BP 32, El Alia, Bab Ezzouar, 16111 Alger, Alg\'{e}rie}
\author{S.~Frullani}
\affiliation{INFN, Sezione Sanit\'a and Istituto Superiore di Sanit\'a, Laboratorio di Fisica, I-00161 Rome, Italy}
\author{C.~Furget}
\affiliation{Laboratoire de Physique Subatomique et de Cosmologie, F-38026 Grenoble, France}
\author{F.~Garibaldi}
\affiliation{INFN, Sezione Sanit\'a and Istituto Superiore di Sanit\'a, Laboratorio di Fisica, I-00161 Rome, Italy}
\author{O.~Gayou}
\affiliation{College of William and Mary, Williamsburg, Virginia 23187, USA}
\author{S.~Gilad}
\affiliation{Massachusetts Institute of Technology, Cambridge, Massachusetts 02139, USA}
\author{R.~Gilman}
\affiliation{Rutgers, The State University of New Jersey, Piscataway, New Jersey 08854, USA}
\affiliation{Thomas Jefferson National Accelerator Facility, Newport News, Virginia 23606, USA}
\author{C.~Glashausser}
\affiliation{Rutgers, The State University of New Jersey, Piscataway, New Jersey 08854, USA}
\author{J.-O.~Hansen}
\affiliation{Thomas Jefferson National Accelerator Facility, Newport News, Virginia 23606, USA}
\author{D.~W.~Higinbotham}
\affiliation{Massachusetts Institute of Technology, Cambridge, Massachusetts 02139, USA}
\affiliation{Thomas Jefferson National Accelerator Facility, Newport News, Virginia 23606, USA}
\author{A.~Hotta}
\affiliation{University of Massachusetts, Amherst, Massachusetts 01003, USA}
\author{B.~Hu}
\affiliation{Hampton University, Hampton, Virginia 23668, USA}
\author{M.~Iodice}
\affiliation{INFN, Sezione Sanit\'a and Istituto Superiore di Sanit\'a, Laboratorio di Fisica, I-00161 Rome, Italy}
\author{R.~Iomni}
\affiliation{INFN, Sezione Sanit\'a and Istituto Superiore di Sanit\'a, Laboratorio di Fisica, I-00161 Rome, Italy}
\author{C.~W.~de~Jager}
\affiliation{Thomas Jefferson National Accelerator Facility, Newport News, Virginia 23606, USA}
\author{X.~Jiang}
\affiliation{Rutgers, The State University of New Jersey, Piscataway, New Jersey 08854, USA}
\author{M.~K.~Jones}
\affiliation{Thomas Jefferson National Accelerator Facility, Newport News, Virginia 23606, USA}
\affiliation{University of Maryland, College Park, Maryland 20742, USA}
\author{J.~J.~Kelly}
\affiliation{University of Maryland, College Park, Maryland 20742, USA}
\author{S.~Kox}
\affiliation{Laboratoire de Physique Subatomique et de Cosmologie, F-38026 Grenoble, France}
\author{M.~Kuss}
\affiliation{Thomas Jefferson National Accelerator Facility, Newport News, Virginia 23606, USA}
\author{J.~M.~Laget}
\affiliation{CEA-Saclay, F-91191 Gif –Sur-Yvette Cedex, France}
\author{R.~De~Leo}
\affiliation{INFN, Sezione di Bari and University of Bari, I-70126 Bari, Italy}
\author{J.~J.~LeRose}
\affiliation{Thomas Jefferson National Accelerator Facility, Newport News, Virginia 23606, USA}
\author{E.~Liatard}
\affiliation{Laboratoire de Physique Subatomique et de Cosmologie, F-38026 Grenoble, France}
\author{R.~Lindgren}
\affiliation{University of Virginia, Charlottesville, Virginia 22901, USA}
\author{N.~Liyanage}
\affiliation{Thomas Jefferson National Accelerator Facility, Newport News, Virginia 23606, USA}
\author{R.~W.~Lourie}
\affiliation{State University of New York at Stony Brook, Stony Brook, New York 11794, USA}
\author{S.~Malov}
\affiliation{Rutgers, The State University of New Jersey, Piscataway, New Jersey 08854, USA}
\author{D.~J.~Margaziotis}
\affiliation{California State University, Los Angeles, Los Angeles, California 90032, USA}
\author{P.~Markowitz}
\affiliation{Florida International University, Miami, Florida 33199, USA}
\author{F.~Merchez}
\affiliation{Laboratoire de Physique Subatomique et de Cosmologie, F-38026 Grenoble, France}
\author{R.~Michaels}
\affiliation{Thomas Jefferson National Accelerator Facility, Newport News, Virginia 23606, USA}
\author{J.~Mitchell}
\affiliation{Thomas Jefferson National Accelerator Facility, Newport News, Virginia 23606, USA}
\author{J.~Mougey}
\affiliation{Laboratoire de Physique Subatomique et de Cosmologie, F-38026 Grenoble, France}
\author{C.~F.~Perdrisat}
\affiliation{College of William and Mary, Williamsburg, Virginia 23187, USA}
\author{V.~A.~Punjabi}
\affiliation{Norfolk State University, Norfolk, Virginia 23504, USA}
\author{G.~Qu\'em\'ener}
\affiliation{Laboratoire de Physique Subatomique et de Cosmologie, F-38026 Grenoble, France}
\author{R.~D.~Ransome}
\affiliation{Rutgers, The State University of New Jersey, Piscataway, New Jersey 08854, USA}
\author{J.-S.~R\'eal}
\affiliation{Laboratoire de Physique Subatomique et de Cosmologie, F-38026 Grenoble, France}
\author{R.~Roch\'e}
\affiliation{Florida State University, Tallahassee, Florida 32306, USA}
\author{F.~Sabati\'e}
\affiliation{Old Dominion University, Norfolk, Virginia 23529, USA}
\author{A.~Saha}
\affiliation{Thomas Jefferson National Accelerator Facility, Newport News, Virginia 23606, USA}
\author{D.~Simon}
\affiliation{Old Dominion University, Norfolk, Virginia 23529, USA}
\author{S.~Strauch}
\affiliation{Rutgers, The State University of New Jersey, Piscataway, New Jersey 08854, USA}
\author{R.~Suleiman}
\affiliation{Massachusetts Institute of Technology, Cambridge, Massachusetts 02139, USA}
\author{T.~Tamae}
\affiliation{Laboratory of Nuclear Science, Tohoku University, Sendai 982-0826, Japan}
\author{J.~A.~Templon}
\affiliation{University of Georgia, Athens, Georgia 30602, USA}
\author{R.~Tieulent}
\affiliation{Laboratoire de Physique Subatomique et de Cosmologie, F-38026 Grenoble, France}
\author{H.~Ueno}
\affiliation{Yamagata University, Kojirakawa-machi 1-4-12, Yamagata 990-8560, Japan}
\author{P.~E.~Ulmer}
\affiliation{Old Dominion University, Norfolk, Virginia 23529, USA}
\author{G.~M.~Urciuoli}
\affiliation{INFN, Sezione Sanit\'a and Istituto Superiore di Sanit\'a, Laboratorio di Fisica, I-00161 Rome, Italy}
\author{E.~Voutier}
\affiliation{Laboratoire de Physique Subatomique et de Cosmologie, F-38026 Grenoble, France}
\author{K.~Wijesooriya}
\affiliation{University of Illinois, Champaign-Urbana, Illinois 61801, USA}
\author{B.~Wojtsekhowski}
\affiliation{Thomas Jefferson National Accelerator Facility, Newport News, Virginia 23606, USA}
 
\collaboration{Jefferson Lab Hall A Collaboration}

\date{\today}

\begin{abstract}

Results of the Jefferson Lab Hall A quasi-elastic {\reaccont}
measurements are presented.
These measurements were performed at fixed transferred momentum and energy, 
$q$ = 1502 MeV/$c$ and $\omega$ = 840 MeV, respectively, 
for missing momenta $p_m$ up to 1 GeV/$c$ and 
missing energies
in the continuum region, up to pion threshold; this kinematic coverage
is much more extensive than that of any previous experiment.
The cross section data are presented 
along with the effective momentum density distribution 
and compared to theoretical models.

\end{abstract}

 
\maketitle

The role of correlations in nuclear structure remains a topic of
primary importance.
On the theoretical side, it is clear that correlations in the nuclear
wave function must exist.
The question is whether correlations can be understood as arising from
the $NN$ force and three-body forces, or whether it will be necessary
to invoke quark degrees of freedom. 
However, the role of correlations in the available experimental data
is often obscure.
Attempts to make more definitive measurements with exclusive $(e,e'p)$ or
$(e,e'NN)$ measurements suffer from reaction mechanism ambiguities;
physics such as meson-exchange currents (MEC), isobar configurations (IC), and
final-state interactions (FSI) mask the effects of correlations.
Better experimental data are needed for a definitive conclusion.

In this letter, we present an attempt to improve the study of correlations,
using the {\reaccont} reaction.
The measurements are in kinematics designed to suppress the effects of
some of the complicating underlying physics and enhance correlations.
In particular, we choose 
high four-momentum transfer $Q^2$ to suppress MEC
and to get a resolution smaller than the nucleon size,
quasifree kinematics $x \approx 1$ to suppress IC,
and the continuum three-body breakup (3bbu) channel.
% to enhance correlations, over the two-body breakup (2bbu) channel, 
%as discussed below.
A recent study \cite{liyanage} of $^{16}$O$(e,e^{\prime}p)X$ demonstrates the
greater kinematic coverage and statistical precision achievable with the current 
generation of accelerator facilities; the results presented here for the
{\reaccont} reaction are by a large
margin the most comprehensive, high resolution, experimental investigation
of the continuum region.

To further motivate the studies presented, in particular the choice
of the 3bbu channel, 
we present two signatures of $NN$ correlations in $^3$He
that one might expect to observe in the $(e,e^{\prime}p)$ reaction,
in the absence of complicating reaction mechanisms.
Consider two correlated nucleons, which in their center 
of mass system have equal and opposite momenta, $\pm\vec{p}_c$,
with higher momenta reflecting smaller separations.
An electron scattering on a proton belonging to such a 
correlated pair inside $^3$He \cite{lala,lolo}
transfers energy $\omega$ and momentum $\vec{q}$.
If the spectator nucleon is at rest, 
the struck proton is ejected with 
momentum $\vec{q}-\vec{p}_c$, while the other nucleon of the pair
moves off with the recoil momentum of the reaction, $\vec{p}_c$. 
The spectator nucleon and the undetected nucleon of the pair 
constitute a recoil system of mass:
\begin{equation}
M^{2}_{r} = \left[ M_{spec}+\sqrt{M^2_N+p^2_c} \right]^2 - p^2_c \, .
\end{equation}
Here $M_N$ is the nucleon mass and $M_{spec}$ is 
the mass of the spectator nucleon.
Thus, in this simple reaction mechanism picture, a signature of correlations
is a peak in the 3bbu cross section as a function of missing energy, $E_m$, 
with the position depending on $p_c$: $E_m = M_r + M_p - M_{^3He}$. 
The peak width reflects the motion of the center of mass with
respect to the spectator nucleon.

The total strength of the correlation peak
as a function of momentum yields a second signature of $NN$ correlations.
While we expect $NN$ correlations to generally be more important 
at missing momenta greater than the Fermi momentum,
the 3bbu strength will be enhanced relative to that for 2bbu as missing 
momentum increases due to a reduced probability for the two undetected 
nucleons to form a bound deuteron at high $p_c$.

\begin{center}
\begin{figure}[tp]
\includegraphics[height=6.pc]{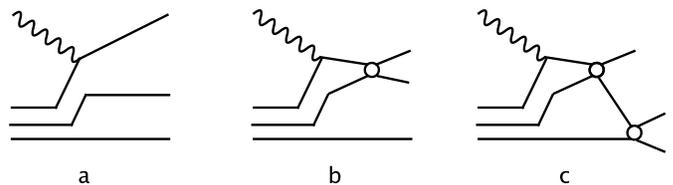}
\caption{Feynman diagrams for a) direct disintegration, 
b) rescattering, and c) rescattering with the spectator nucleon. }
\label{fig:feynman}
\end{figure}
\end{center}

An apparent correlation peak in the continuum was observed 
with limited statistics and limited kinematic range
for the first time at Saclay \cite{March}, on $^3$He,
and subsequently on $^4$He \cite{LeGoff}.
Simply interpreting the data as evidence for correlations,
as suggested by the simple reaction mechanism picture of
Fig.~\ref{fig:feynman}a, neglects the complications
of MEC, IC, and FSI.
In particular, the derivation of the peak position is purely kinematic, 
and the peak can appear as long as the electrodisintegration 
involves two active nucleons plus spectator nucleons.
Thus, when the two nucleons in the active pair rescatter, 
Fig.~\ref{fig:feynman}b, the position and width of the peak do not 
change, and the simple picture remains valid,
but one measures the transition between a correlated pair in the ground 
state and a correlated pair in the continuum.
But when one of the nucleons of the active pair re-interacts 
with the spectator third nucleon, Fig.~\ref{fig:feynman}c, 
the position, shape, and amplitude of the peak might all be affected.
Rescattering also makes the measured missing momentum $p_m$ different
from the correlation momentum $p_c$, so it is clear then that the observed
momentum distribution does not simply reflect the nuclear wave function --
it is not an actual density.
The lack of an apparent correlation peak in 
continuum $(e,e^{\prime}p)$ on heavier nuclei, e.g., $^{16}$O \cite{liyanage},
is believed to result from the small probability of having only two
active nucleons when the number of spectators gets large.

In Thomas Jefferson National Accelerator Facility (JLab) 
Hall A experiment E89-044 \cite{Epst},
we studied the {\reacboth} reaction in the
quasi-elastic region at transferred 3-momentum $|\vec{q}\,|$ = 1502 MeV/$c$ 
and energy $\omega$ = 840 MeV, so $Q^2$ = 1.55 GeV$^2$.
This paper reports the results of measurements 
in perpendicular kinematics with Bjorken $x$ = 0.98, 
near the top of the quasifree peak.
Protons were detected at several angles relative to $\vec{q}$,
corresponding to missing momenta $p_m$ of 0 - 1 GeV/$c$.
Results of the {\reactwobbu} reaction channel from 
this experiment were reported in \cite{Rvac};
here we focus on the continuum {\reaccont} channel, $E_m$ $>$ 7.72 MeV.

A continuous, $\approx$120 $\mu$A, electron beam was scattered from $^3$He in a 
10 cm diameter cylindrical cell, mounted with the beam
passing through the center of the target perpendicular to the symmetry axis.
The $^{3}$He target density was $\approx$ 0.072 g/cm$^3$ \cite{Rvac}. 
The scattered electrons and knocked-out protons were detected
in the two High-Resolution Spectrometers (HRS$_e$ and HRS$_h$).
Details of the Hall A experimental setup are given in
\cite{halla}; see \cite{fbphd} for further details of this experiment.

\begin{table}[b]
\begin{center}
\caption{Proton spectrometer kinematic settings.}
\begin{tabular}{|c|c|c|}
\hline\hline
 \hspace*{9.9mm}$p_m$\hspace*{9.9mm} & 
 \hspace*{9.9mm}$P_p$\hspace*{9.9mm} &
 \hspace*{9.9mm}$\theta_{p}$\hspace*{9.9mm} \\
 (MeV/$c$) & (MeV/$c$) & ($^\circ$) \\
\hline
150 & 1493 & 54.04 \\
300 & 1472 & 59.83 \\
425 & 1444 & 64.76 \\
550 & 1406 & 69.80 \\
750 & 1327 & 78.28 \\
1000 & 1171 & 89.95 \\
\hline\hline
\end{tabular}
\label{tab:ExpKin}
\end{center}
\end{table}

Throughout the experiment, singles {\reacqe} quasi-elastic scattering data,
measured simultaneously with coincidence {\reacboth}, provided a
continuous monitor of both luminosity and beam energy. 
The absolute normalization of our data was determined by comparing
measurements of elastic scattering data to world data \cite{Amroun}. 
We measured the {\reacx} cross section at three beam
energies, keeping $\vert\vec q\thinspace\vert$ and $\omega$ fixed
in order to separate response functions and understand systematic
uncertainties. 
The data reported in this paper were all obtained at
a beam energy of 4806 MeV.

The missing energy resolution, about 1 MeV ($\sigma$),
is less than the 2.23 MeV separation between the {\reactwobbu} 
peak and the threshold for the {\reaccont} breakup channels.
The radiative corrections to the measured cross sections were
performed by using the code MCEEP~\cite{ulmer2}. 
The radiative tail is simulated and folded into the ($E_m$, $p_m$)
space based on the prescription of Borie and Drechsel~\cite{borie}. 
The radiative corrections in the continuum amount to ${\rm 10-20\%}$ 
of the cross section.
In particular, the radiative corrections remove the tail of the 2bbu
process from the 3bbu data, allowing a clear separation of the channels.
An exception is for low missing momentum, below 100 MeV/$c$,
where the 3bbu strength is less than the strength of the 
radiative tail of the much stronger 2bbu peak.

\begin{figure}[tp!]
\begin{center}
\includegraphics[height=43.0pc]{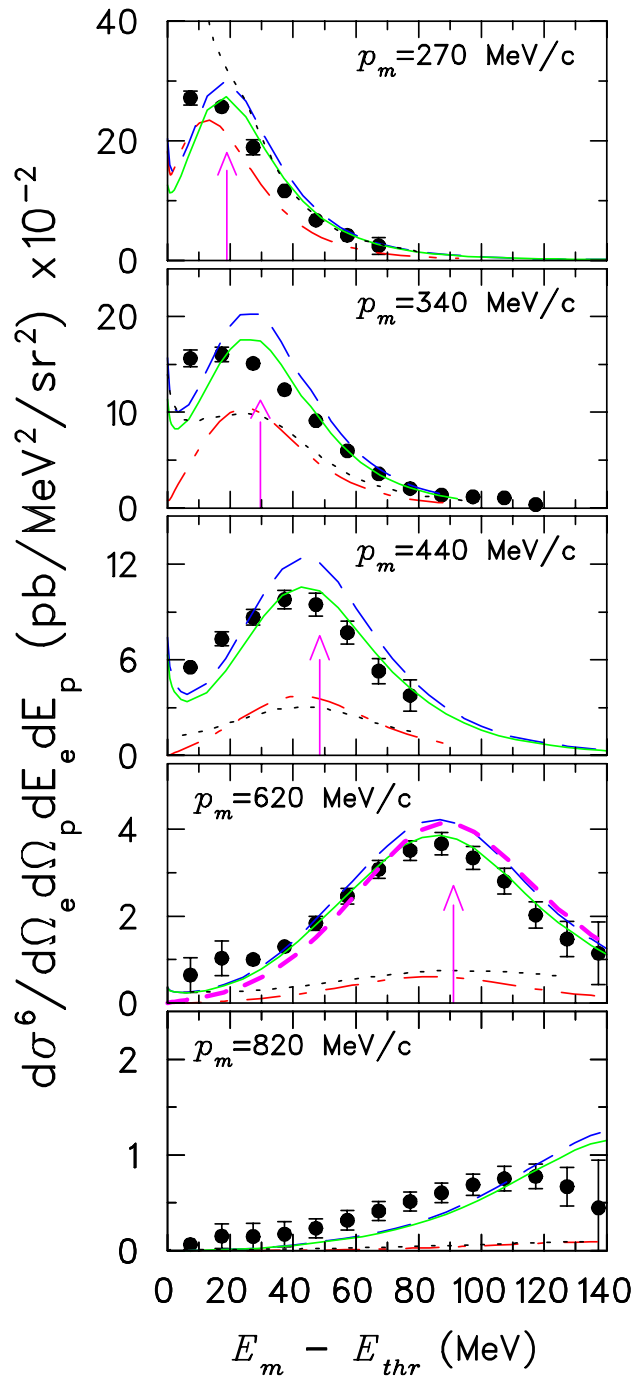}
\caption{(color online). Cross-section results for the 
{\reaccont} reaction versus missing energy $E_m$.
The vertical arrow gives the peak position expected for disintegration
of correlated pairs.
The dotted curve presents a PWIA calculation using Salme's spectral 
function and $\sigma_{cc1}$ electron-proton off-shell cross section.
Other curves are recent theoretical predictions of J. M. Laget \cite{Lag4} from 
the PWIA (dash dot) to PWIA + FSI (long dash) to full calculation (solid), 
including meson exchange current and final state interactions.
In the 620 MeV/$c$ panel, the additional short dash curve is a calculation
with PWIA + FSI only within the correlated pair.
}
\label{fig:crosssections}
\end{center}
\end{figure}

Table~\ref{tab:ExpKin} shows the central proton spectrometer settings for
the experimental kinematics presented in this paper.  
The data taken at these settings are grouped into numerous ($E_m$, $p_m$) bins
for presentation; Fig.~\ref{fig:crosssections} shows the 
cross sections
corrected for radiative processes as a function of missing energy for 
several selected bins.
The energy scale in the horizontal axis has been shifted in these plots 
so that the 3bbu channel starts at 0.
As $p_m$ increases, we can see that the broad peak in the cross
section moves to higher missing energies.
The arrow in the figure indicates where one would expect the peak in 
the cross section due to disintegration
processes involving two active nucleons plus a spectator; the expected 
peak position for $p_m$ $=$ 820 MeV/$c$ is just off scale, at $E_m$ $\approx$
145 MeV.
The large peak in the data roughly aligned with the arrow 
suggests that two-nucleon disintegration processes are dominant.

Several calculations are presented in Fig.~\ref{fig:crosssections}.
The simplest calculation is a plane-wave impulse approximation
(PWIA) calculation using Salme's spectral 
function \cite{Kiev} and the $\sigma_{cc1}$ electron-proton off-shell 
cross section \cite{defor}.
Also shown in Fig.~\ref{fig:crosssections} are the results of
microscopic calculations of the continuum cross section 
by J. M. Laget \cite{Lag1}, including a PWIA calculation with 
correlations but no FSI,
and successive implementation of various interaction effects.
The calculation is based on a diagrammatic expansion of the reaction amplitude,
up to and including two loops \cite{Lag2}. 
Both single and double ${\rm NN}$ scattering, as well as meson exchange
($\pi$ and $\rho$)  
and $\Delta$ formation are included.  
The bound-state wave function is a solution of the Faddeev 
equation used by the Hannover group \cite{Hadj} for
the Paris potential \cite{Laco}. 
Nucleon and meson
propagators are relativistic and no angular approximations
(Glauber) have been made in the various loop integrals. 
The FSI in these calculations use a global parameterization of the $NN$
scattering amplitude, obtained from experiments at LANL, SATURNE and
COSY~\cite{Lag4}.
Further details of the model can be found in \cite{Lag3}.

Figure \ref{fig:crosssections} shows that the calculated cross sections
exhibit a correlation peak that is dominant at low $p_m$, 
but that FSI strongly enhance the cross section at large $p_m$.
The calculations indicate the
FSI are mainly between the two active nucleons -- Fig.~\ref{fig:feynman}b.
The additional calculation
included in the 620 MeV/$c$ panel of Fig.~\ref{fig:crosssections}
has FSI with the spectator nucleon -- Fig.~\ref{fig:feynman}c -- turned off.
Neither the shape nor magnitude of the peak is much affected.
This result indicates that triple rescattering is negligible.
MEC effects are also small.

\begin{center}
\begin{figure}[b]
\includegraphics[height=18.pc]{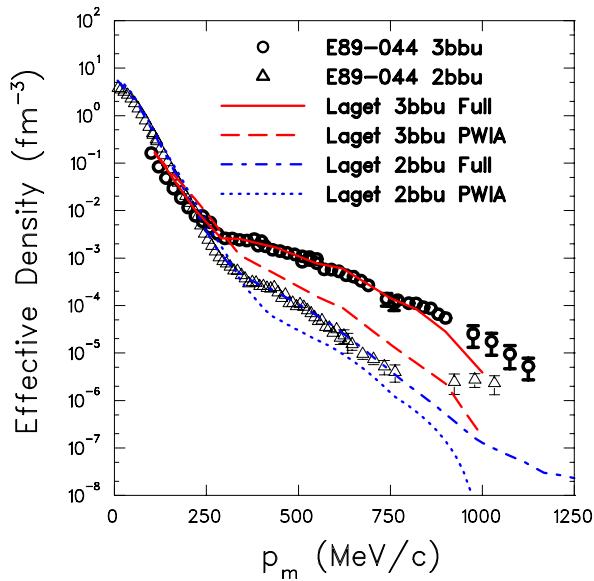}
\caption{(color online). 
Proton effective momentum density distributions in $^3$He 
extracted from {\reaccont} (open black circles) and {\reactwobbu} 
(open black triangles), compared to calculations from Laget \cite{Lag4}.
The 3bbu integration covers $E_M$ from threshold to 140 MeV.}
\label{fig:Density}
\end{figure}
\end{center}

To obtain the total 3bbu strength, and to facilitate comparison
to the 2bbu,
we divided the cross section by the elementary off-shell 
electron-proton cross section $\sigma_{ep}$ \cite{defor},
multiplied by a kinematic factor $K$, 
and integrated over missing energy to obtain the
effective momentum density distribution: 
\begin{equation}
\eta{(p_{m})} = \int\bigl({\frac{d^6\sigma}{dE_edE_pd\Omega_ed\Omega_p}}
\, / \, K \sigma_{ep} \bigr) \,\ dE_{m}.
\label{eq:emint}
\end{equation}
Figure~\ref{fig:Density} shows the distribution obtained.
Uncertainties from missing tails of the 3bbu peak, within 
this integration range, due to
limited experimental acceptance are negligible on the scale of
Fig.~\ref{fig:Density}.
The 3bbu distribution tends to have a much larger relative 
strength for high missing momentum than does the 2bbu distribution
- the ratio of 3bbu to 2bbu strength increases with $p_m$ by about 
three orders of magnitude, from about 100 to 800 MeV/$c$.
An increase of the relative strength with $p_m$ is consistent 
with what is expected from correlations,
as described in the simple picture in the introduction, but
we already know from the discussion of Fig.~\ref{fig:crosssections} 
that FSI are important.

The PWIA curves in Fig.~\ref{fig:Density} show an order of magnitude 
enhancement of the 3bbu over the 2bbu at high missing momentum.
The two-body correlations are more clearly seen in 3bbu than 
in the 2bbu since the available phase space is reduced when two nucleons 
are forced to form the deuteron.
The differences between the PWIA calculations %and the data 
and full 
calculations further indicate the greater importance of final-state 
interactions in the 3bbu. 
Thus, the larger FSI in the 3bbu mask the larger role of correlations.
The generally good agreement of the full calculations and the data shown
in Figs.~\ref{fig:crosssections} and \ref{fig:Density}
relies mainly on the interplay of correlations and final-state interactions, 
and indicates no need for physics beyond that already present in a modern
conventional nuclear physics model.
The conclusions presented here have been confirmed by subsequent, independent
calculations \cite{cioffi}.

The conclusions described above might appear to be no longer valid for
$p_m$ $\approx$ 1 GeV/$c$
as the magnitude of the 3bbu appears to fall towards that of the 2bbu.
However, the center of the 3bbu correlation peak moves
outside of the integration range at $p_m$ $\approx$ 800 MeV/$c$, as
shown in Fig.~\ref{fig:crosssections}.
Thus, the experimental integration only includes a fraction of the
3bbu strength at large $p_m$.
A crude correction to account for the missing strength, 
using the fraction of strength of the full calculation of Laget 
in the region $E_m$ $<$ 140 MeV, 
causes the 3bbu
distribution to roughly flatten out, starting near 750 MeV/$c$, at a level
nearly two orders of magnitude greater than that of the 2bbu.
The large correction also leads to our stopping the calculation
at 1 GeV/$c$; the comparison between data and theory is no longer meaningful
when only a small fraction of the tail of the distribution is considered.
Given these data along with the theoretical calculations, it remains
fair to conclude that the correlations in the wave function 
preferentially lead to the 3bbu channel, and that the reaction
mechanism is reasonably well understood in a modern, conventional
nuclear physics model.

The comparison of the data of this experiment with the existing
calculation suggests that the region near 300 MeV/$c$
might prove to be the most enlightening with respect to the
role of correlations.
Here the full and PWIA curves are very similar
to each other and to the data, and in the theory 
the correlation peak dominates the cross section.
Separated response functions, which are possible with data from the other
kinematics of this experiment, can provide us with more complete tests of 
the theory.

In summary, results for the cross section at constant $\vec{q}$ and
$\omega$ have been presented for the {\reaccont} reaction channel. 
The experimental data are both much higher in statistics and more extensive
in kinematic coverage than any previous measurement.
Model calculations are in good agreement with the data.
We believe these are benchmark data which will serve to stimulate
additional independent calculations, and help to define
the role of correlations in nuclear structure.

We acknowledge the outstanding support of the staff of the Accelerator
and Physics Divisions at JLab that made this experiment
successful. 
This work was supported in part by
the U.S. Department of Energy contract DE-AC05-84ER40150 Modification No. M175
under which the Southeastern Universities Research Association
(SURA) operates the Thomas Jefferson National Accelerator Facility, 
other Department of Energy contracts, 
the U.S. National Science Foundation, 
the Italian Istituto Nazionale di Fisica Nucleare (INFN), 
the French Atomic Energy Commission and National Center of Scientific Research,
the Natural Sciences and Engineering Research Council of Canada, and
Grant-in-Aid for Scientific Research (KAKENHI) (No.\ 14540239) from 
the Japan Society for Promotion of Science (JSPS).

\end{document}